\documentclass[aps,showpacs,twocolumn]{revtex4}
\usepackage{amsfonts}
\usepackage{amsmath}
\usepackage{amssymb}
\usepackage{graphicx}
\usepackage{bm}

\input{tcilatex}

\begin{document}

\title{Charging of heated colloidal particles using the electrolyte Seebeck
effect}
\author{Arghya Majee and Alois W\"{u}rger}
\affiliation{Laboratoire Ondes et Mati\`{e}re d'Aquitaine, Universit\'{e} de Bordeaux \&
CNRS, 351 cours de la Lib\'{e}ration, 33405 Talence, France}

\begin{abstract}
We propose a novel actuation mechanism for colloids, which is based on the
Seebeck effect of the electrolyte solution: Laser heating of a nonionic
particle accumulates in its vicinity a net charge $Q$, which is proportional
to the excess temperature at the particle surface. The corresponding
long-range thermoelectric field $E\propto 1/r^{2}$ provides a tool for
controlled interactions with nearby beads or with additional molecular
solutes. An external field $E_{\text{ext}}$ drags the thermocharged particle
at a velocity that depends on its size and absorption properties; the latter
point could be particularly relevant for separating carbon nanotubes
according to their electronic bandstructure.\textbf{\ }
\end{abstract}

\maketitle

Selective transport and controlled pattern formation are of fundamental
interest in microfluidics and biotechnology \cite{Squ05}. Particle focussing
devices \cite{Abe08,Pri08,Pal10} and macromolecular traps \cite%
{Duh06,Jia09,Wie10} have been designed by applying chemical or thermal
gradients. In \textquotedblleft active colloids\textquotedblright\ there is
no external symmetry breaking field: Thermodynamic forces arise from an
embarked chemical reactor \cite{Gol05,How07} or from non-uniform laser
heating of Janus particles \cite{Jia10}. In both cases the colloid
self-propels in an anisotropic environment that is created by the
concentration or temperature variation along its surface.\ The interplay of
self-propulsion and Brownian motion leads to a complex diffusion behavior 
\cite{Gol05,How07,Gol09,Jia10}.

Locally modifying material properties by heating a single nanoparticle or
molecule in a focussed laser beam is by now a standard technique. The
temperature dependent refractive index was used for the photothermal
detection of a single non-fluorescent chromophore \cite{Gai10}. Heating a
spherical nanoparticle induces a radial temperature profile in the
surrounding fluid and, because of the viscosity change, an enhancement of
the Einstein coefficient \cite{Rin10}. The non-uniform laser absorption of
half-metal coated particles leads to a temperature variation along its
surface; the resulting self-propulsion adds a ballistic velocity component
and thus increases the effective mean-square displacement \cite{Jia10}.%

\begin{figure}
\includegraphics[width=\columnwidth]{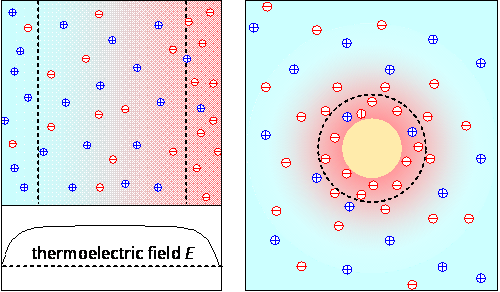}
\caption{Left panel: Seebeck effect in
an electrolyte solution that is cooled at the left side and heated at the
right. We show the case of a positive Seebeck coefficient $S$, where
cations and anions accumulate at the cold and hot boundaries, respectively.
These charged layers have a thickness of about one Debye length (dashed
lines). The corresponding thermoelectric field $E$ is constant in the bulk
and vanishes at the boundaries \protect\cite{Maj11}. Right panel: Seebeck
effect in the vicinity of a hot particle with excess temperature $\protect%
\delta T$. Due to the radial temperature gradient, a net charge $Q$
accumulates within one Debye length from the particle surface. The charge
density and the radial electric field are shown in Fig. 2 below. The
counterions are at the vessel boundary.}
\end{figure}

In this Letter, we point out that heating confers a net charge on colloidal
particles, and how this electrolyte Seebeck effect in the vicinity of a
heated bead can be used for selective transport and controlled interactions.
After a brief reminder of the well-known case of a constant temperature
gradient, we derive the expressions for the thermocharge and the electric
field induced by a spherical particle with excess temperature $\delta T$. As
possible applications, we then discuss the thermoelectric pair potential,
aggregation or depletion of a molecular solute in the vicinity of a hot
particle, and colloidal separation through velocity differentiation.

The response of a salt solution to a constant thermal gradient $\nabla T$ is
illustrated in the left panel of Fig.\ 1. Because of their temperature
dependent solvation energy, positive and negative salt ions migrate along
the gradient. In general one of the species moves more rapidly, resulting in
a thermopotential between the cold and hot boundaries of the sample \cite%
{Eas28} and a macroscopic electric field $E=S\nabla T$, which is
proportional to the thermal gradient and to the Seebeck coefficient $S$ \cite%
{Aga89}. The bulk solution is neutral; yet opposite charges accumulate at
the boundaries and screen the electric field in a layer of one Debye length.
In the last years it has become clear that charged colloids in a temperature
gradient are a sensitive probe to the Seebeck effect of the electrolyte
solution \cite{Put05,Vig10}: The field $E$ and thus the colloidal velocity
depend strongly on the salt composition and are particularly important in
the presence of molecular ions containing hydrogen \cite{Wue08,Wue10,Maj11};
this thermo-electrophoretic driving has been confirmed for SDS micelles in a
NaCl$_{1-x}$OH$_{x}$ solution, where a change of sign of the drift velocity
has been observed upon varying the parameter $x$ \cite{Vig10}.

\textit{Thermocharge of a hot colloid. }Now we consider the Seebeck effect
in the vicinity of a heated particle, as shown in the right panel of Fig.\
1. The qualitative features are readily obtained by wrapping the hot
boundary of the one-dimensional case (left panel) onto a sphere of radius $a$%
.\ Its excess temperature $\delta T$ results in a thermal gradient 
\begin{equation}
\nabla T=-\frac{\delta Ta}{r^{2}}  \label{1}
\end{equation}%
and, at distances well beyond the Debye length, in a thermoelectric field $%
E=S\nabla T$. Its complete expression, in particular close to the particle
surface, is obtained from the stationary electro-osmotic equations for the
ion currents

\begin{equation}
J_{\pm }=-D_{\pm }\left( \nabla n_{\pm }\mp n_{\pm }\frac{eE}{k_{B}T}%
+2n_{\pm }\alpha _{\pm }\frac{\nabla T}{T}\right) ,  \label{3}
\end{equation}%
which comprise normal diffusion with coefficients $D_{\pm }$,
electrophoresis with the H\"{u}ckel mobility for monovalent ions, and
thermal diffusion with parameters $\alpha _{\pm }$. The latter are reduced
values of the ionic Soret coefficient, introduced by Eastman as a measure
for the electromotive force of an electrolyte thermocouple \cite{Eas28};
experimental values are found in Refs. \cite{Aga89,Sok06,Bon11}.

\begin{figure}[b]
\includegraphics[width=\columnwidth]{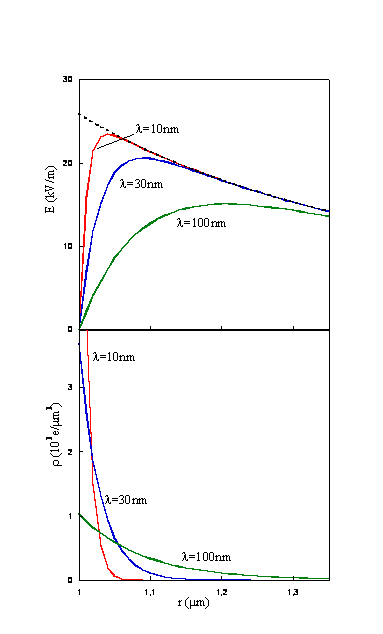}
\caption{Thermoelectric field $E$ and charge
density $\protect\rho $ as a function of the distance $r$ from the particle
centre for different values of the Debye length $\protect\lambda $ and fixed
particle size $a=1\protect\mu $m. We have used the parameters $\protect%
\alpha _{+}-\protect\alpha _{-}=-10$ and $\protect\delta T=30$\ K. The
dashed line gives the bulk law $E=S\protect\nabla T$.}
\end{figure}

The thermoelectric field $E$ and the charge density $\rho =e(n_{+}-n_{-})$
are obtained from the steady-state condition $J_{\pm }=0$ and Gauss' law $%
\mathrm{div}E=\rho /\varepsilon $. Linearizing the currents in the small
gradients and solving the coupled differential equations for $E$ and $\rho $%
, we obtain the radial steady-state thermoelectric field 
\begin{equation}
E=S\nabla T\left( 1-\frac{r+\lambda }{a+\lambda }e^{(a-r)/\lambda }\right) ,
\label{2}
\end{equation}%
with the Debye length $\lambda $ and the Seebeck coefficient $S=(\alpha
_{+}-\alpha _{-})k_{B}/e$. For a detailed calculation see Ref. \cite{SI}.
With (\ref{1}) the field $E$ is zero on the particle surface and at
infinity, as required by electrostatic boundary conditions.\ At distances
well beyond $\lambda $ the exponential factor vanishes; the remaining
long-range contribution $E=S\nabla T$ varies with the inverse square of the
distance $r$. From Gauss' law one obtains the charge density \cite{SI} 
\begin{equation*}
\rho =\frac{Qe^{(a-r)/\lambda }}{4\pi (a+\lambda )\lambda r},
\end{equation*}%
which is concentrated within about one Debye length from the particle
surface. Fig.\ 2 illustrates the variation of $E$ and $\rho $ with distance
for different values of $\lambda $; the former is long-range whereas the
latter decays exponentially. The net thermocharge carried by an otherwise
non-ionic particle, 
\begin{equation}
Q=-e(\alpha _{+}-\alpha _{-})\frac{a}{\ell _{B}}\frac{\delta T}{T},
\label{4}
\end{equation}%
depends on its radius in units of the Bjerrum length $\ell _{B}=7$ \AA , the
ratio of excess and absolute temperature, and the reduced Seebeck parameter $%
\alpha _{+}-\alpha _{-}$ \cite{Wue10}.

In physical terms the charge $Q$ arises from the difference in
thermo-osmotic pressure of positive and negative ions, which in turn is
related to the ionic Soret parameters $\alpha _{\pm }/T$ \cite{Eas28}. For $%
\alpha _{+}>\alpha _{-}$ the anions show thermal diffusion toward higher
temperature, thus accumulating a negative charge at the particle surface.
The corresponding cations are located at $r\rightarrow \infty $; in the case
of a spherical sample container of radius $R$, the corresponding surface
charge density $Q/4\pi R^{2}$ is very small. Numerical values for the
Seebeck coefficient of several electrolytes are given in Table 1. For small
ions the numbers $\alpha _{\pm }$ are of the order of unity; higher values
occur for molecules containing hydrogen. For a 100 nm-bead in NaOH or HCl
solution with $\delta T=30$ K, one finds that $Q$ corresponds to about 40
elementary charges; still higher values occur for protonated salts. 
\begin{table}[tbp]
\caption{Seebeck coefficient $S$ for NaCl, HCl and NaOH in aqueous solution 
\protect\cite{Aga89,Sok06}, and for tetrabutylammonium nitrate (TBAN) in
water (w) and dodecanol (d) \protect\cite{Bon11}. For comparison, $S$ of
simple metals is of the order of a few $\protect\mu $V/K. The Seebeck
coefficient is related to Eastman's ionic entropy of transfer $2k_{B}\protect%
\alpha _{\pm }$ through $S=(k_{B}/e)(\protect\alpha _{+}-\protect\alpha %
_{-}) $ \protect\cite{Wue10}. Experimental values for various ions are given
in Refs. \protect\cite{Eas28,Aga89,Sok06,Bon11}. }%
\begin{tabular}{|l|c|c|c|c|c|}
\hline
Salt/solvent & NaCl/w & NaOH/w & HCl/w & TBAN/w & TBAN/d \\ \hline
$S$ (mV/K) & $0.05$ & $-0.22$ & $0.21$ & $1.0$ & $7.2$ \\ \hline
$\alpha _{+}-\alpha _{-}$ & $0.6$ & $-2.7$ & $2.6$ & $12$ & $86$ \\ \hline
\end{tabular}%
\end{table}

Equations (\ref{2}) and (\ref{4}) are the main formal results of this paper.
In the remainder we discuss how the thermocharge allows to actuate colloidal
motility and interactions, and how the thermoelectric field can be used for
locally accumulating or depleting an additional charged molecular solute. As
an overall feature we estimate the thermoelectric response time. Because of
the fast equilibration of heat flow and temperature, thermocharging occurs
on the time scale of thermal diffusion of salt ions over one Debye length.
With the above parameters one finds a relaxation time of the order of
microseconds. Thus on the scale of colloidal motion, thermocharging is an
almost instantaneous process.

\textit{Colloid-colloid forces. }We start with the electric force $QE$
between two hot particles at a distance $R$.\ Assuming $R\gg \lambda $ and
using the definition of the Bjerrum length, we find 
\begin{equation}
F=\frac{Q^{2}}{4\pi \varepsilon R^{2}}.  \label{8}
\end{equation}%
Thus in an electrolyte with finite Seebeck coefficient, heating disperses
colloidal aggregates and strongly affects collective effects due to
thermophoretic or hydrodynamic interactions \cite{Gol11}. So far we have
considered non-ionic colloids. A particle carrying a proper charge $Q_{p}$
gives rise to an additional electric field $E_{p}=Q_{p}e^{-(r-a)/\lambda
}/(4\pi \varepsilon \lambda r)$. Depending on the sign of $Q$ and $Q_{p}$,\
the superposition $E+E_{p}$ shows a complex spatial variation; note that $%
E_{p}$ is screened whereas $E$ is not.

\textit{Thermo-electrophoresis. }Thermocharging provides a unique tool for
creating a radial electric field in an electrolyte solution. For a
micron-size bead with an excess temperature $\delta T=30$ K, the field $E$
may attain $10^{4}$ V/m in its immediate vicinity, and a few V/m at a
distance of 100 microns. The electrophoretic velocity of a molecular solute
with zeta potential $\zeta $, 
\begin{equation}
u=\frac{2}{3}\frac{\varepsilon \zeta }{\eta }E,  \label{7}
\end{equation}%
varies between 10 $\mu $m/s and 10 nm/s. Depending on the sign of the zeta
potential $\zeta $ and of the Seebeck coefficient, molecular ions are
attracted or repelled by the thermocharge. As illustrated in Fig.\ 3a., this
can be used for accumulating or depleting a molecular solute in the vicinity
of the particle.

More complex patterns are realized by superposing $E$ with the screened
field $E_{p}$ of a proper charge. In addition to thermo-electrophoresis (\ref%
{7}), the radial temperature profile results in thermal diffusion of the
solute molecules, due to both double-layer and dispersion forces \cite%
{Pia08,Wue10}. Finally we mention that the thermoelectric field $E$ of Janus
particles comprises a strongly anisotropic short-range component.

\textit{Selective transport.} Sorting molecular or colloidal solutes by size
is of interest for various applications. The sedimentation potential being
rather ineffective for submicron particles, common methods are based on
electrophoresis or on motion driven by thermodynamic forces. Since the
free-solution mobilities are in general independent of size and molecular
weight, velocity differentiation is achieved only after adding a molecular
solute as in gel electrophoresis \cite{Sou87,Vio00}, or by spatial flow or
field modulation \cite{Han00,Hua04,Zhe03}.

\begin{figure}[b]
\includegraphics[width=\columnwidth]{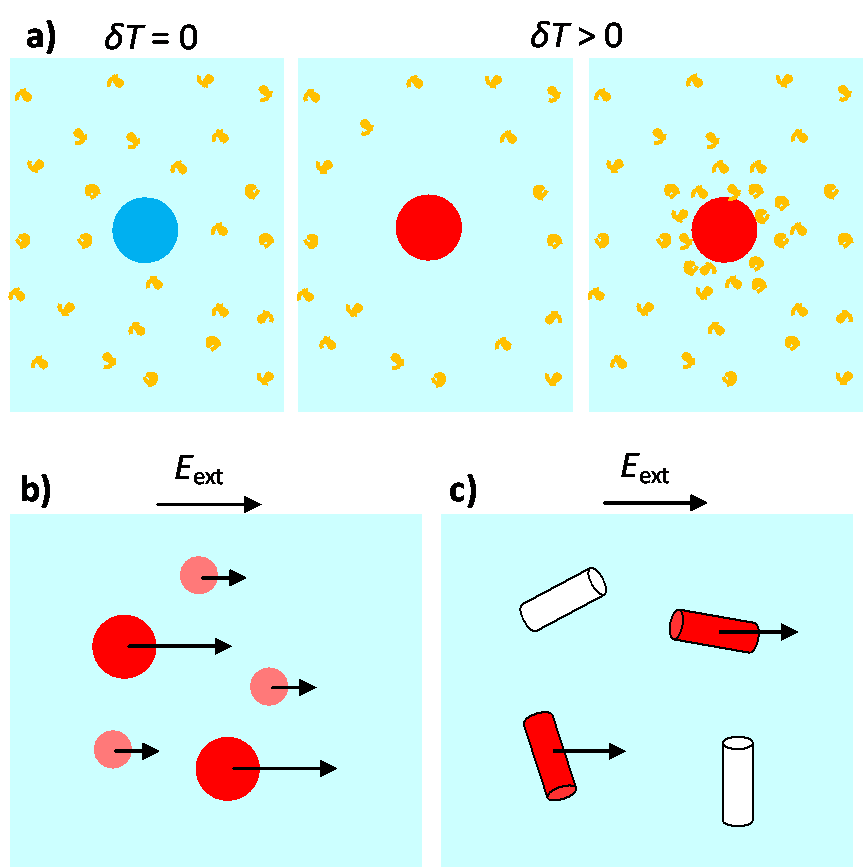}
\caption{Pattern formation and selective transport due to thermocharging.
a) The thermoelectric field of a hot particle induces electrophoretic motion
(\protect\ref{7}) of a charged molecular solute. Depending on the sign of
the electrolyte Seebeck coefficient $S$ and the molecular zeta potential $%
\protect\zeta $, the colloidal thermocharge results in depletion or
accumulation of the solute. b) In an external electric field $E_{\text{ext}}$%
, heated particles with excess temperature $\protect\delta T$ and
thermocharge $Q$, acquire a size-dependent velocity (\protect\ref{6}). c)
Because of their different optical absorption properties, metallic and
semiconducting carbon nanotubes differ in their thermocharge and in their
response to an electric field.}
\end{figure}

Here we show that thermocharging in the presence of an applied electric
field $E_{\text{ext}}$, provides an efficient means for separating particles
by size. The force density $\rho E_{\text{ext}}$ exerted by the external
field on the charged fluid results in a drift velocity $u_{\text{ext}}$ of
the particle; solving the stationary Stokes equation one finds \cite{SI} 
\begin{equation}
u_{\text{ext}}=\frac{QE_{\text{ext}}}{6\pi \eta (a+\lambda )}.  \label{6}
\end{equation}%
Note the explicit dependence on the particle size, in contrast to the
Helmholtz-Smoluchowski mobility. In view of the thermocharge (\ref{4}), the
most interesting dependencies arise from the excess temperature $\delta T$.
Assuming a constant volume absorption coefficient $\beta $, one finds that
the excess temperature varies with the square of the radius,%
\begin{equation}
\delta T=\frac{a^{2}\beta I}{3\kappa },  \label{5}
\end{equation}%
where $I$ is the laser intensity and $\kappa $ the thermal conductivity of
the solvent.

According to (\ref{5}) the excess temperture increases with the square of
the bead size; thus the drift velocity $u_{\text{ext}}$ varies with the
particle surface in the H\"{u}ckel limit ($a<\lambda $) and with its volume
in the case $a>\lambda $, as illustrated in Fig.\ 3b. As an
order-of-magnitude estimate, heating the beads by $\delta T=30$ K and
applying a field $E_{\text{app}}\sim 10^{4}$ V/m results in a velocity of
about 10 $\mu $m/s. The above argument holds true for non-spherical solute
particles, albeit with different geometrical factors. The excess temperature
of metal-coated polystyrene beads is linear in the radius. For polymers the
charge $Q$ is proportional to the chain length or number $N$ of monomers,
whereas the friction coefficient varies with the gyration radius $R\propto
N^{\nu }$, resulting in a velocity $u\propto N^{1-\nu }$. In aqueous
solution most colloids carry a proper charge with surface potential $\zeta
_{p}$, resulting in an additional velocity $u_{p}=(\varepsilon \zeta
_{p}/\eta )E_{\text{ext}}$. Still, the thermocharge leads to a significant
dispersion of the total velocity $u_{\text{ext}}+u_{p}$.

We conclude with a possible application to the separation of carbon
nanotubes by their wrapping structure \cite{Zhe03,Arn06}. The electronic and
optical properties of single-wall nanotubes depend crucially on their
\textquotedblleft chiral vector\textquotedblright\ ($n,m$) that describes
the orientation of the graphene structure with respect to the tube axis.
Depending on these indices, one has metallic or semiconducting nanotubes
with a characteristic bandstructure and a particular optical spectrum. The
excess temperature $\delta T=a\bar{\beta}I/\kappa $ of a nanotube depends on
its radius $a$ and the absorption per unit area $\bar{\beta}$ of its
graphene sheet, and so does the drift velocity $u_{\text{ext}}$. As
illustrated in Fig. 3c, by chosing an appropriate laser wave length, one
could selectively heat nanotubes with a given chiral vector and separate
them through thermocharge electrophoresis.

\end{document}